\documentstyle[epsf]{mn} 

\footnotesize
\newdimen\minuswidth    
\setbox0=\hbox{$-$}
\minuswidth=\wd0
\catcode`@=\active
\def@{\kern\minuswidth}
\newdimen\digitwidth    
\setbox0=\hbox{\rm0}
\digitwidth=\wd0
\catcode`!=\active
\def!{\kern\digitwidth}
\normalsize

\def\plotone#1{\centering \leavevmode
\epsfxsize=\columnwidth \epsfbox{#1}}

\title{Radio Pulsars in Terzan 5}

\author[A. G. Lyne et al.]
{
A. G. Lyne$^1$, 
S. H. Mankelow$^1$,
J. F. Bell$^{2,1}$,
R. N. Manchester$^2$ \\
$^1$ University of Manchester, Jodrell Bank Observatory,
 Macclesfield, Cheshire SK11 9DL, UK\\
$^2$ Australia Telescope National Facility, CSIRO, P.O.Box 76, Epping NSW
 1710, Australia\\
}

\begin{document}
 
\def\lapp{\ifmmode\stackrel{<}{_{\sim}}\else$\stackrel{<}{_{\sim}}$\fi}
\def\gapp{\ifmmode\stackrel{>}{_{\sim}}\else$\stackrel{>}{_{\sim}}$\fi}

\maketitle
\newcommand{\setthebls}{
}
\setthebls

\begin{abstract}
We report on searches of the globular cluster Terzan 5 for low
luminosity and accelerated radio pulsars using the 64-m Parkes
radio telescope.  One new millisecond pulsar, designated
PSR~J1748$-$2446C, was discovered, having a period of 8.44~ms.  Timing
measurements using the 76-m Lovell radio telescope at Jodrell Bank show
that it is a solitary pulsar and lies close to the core of the
cluster. We also present the results of timing measurements which show
that the longer-period pulsar PSR~J1748$-$2444 (formerly known as
PSR~B1744$-$24B) lies 10 arcmin from the core of the cluster and
is unlikely to be associated with the cluster.  We conclude that there
are further pulsars to be detected in the cluster.
\end{abstract}
 
 
\section{INTRODUCTION}

Globular clusters were proposed as good hunting grounds for rapidly
rotating pulsars because they were known to contain a relatively high
proportion of potential progenitor systems such as low mass X-ray
binaries \cite{hhb85}. Since the first globular cluster pulsar was
found in M28 \cite{lbm+87}, several dozen have been found in a number of
clusters \cite{lyn95}. This paved the way for new understanding of the
formation and evolution of millisecond pulsars \cite{ka96}, as well as
providing opportunities for other applications such as measuring
cluster gravitational potentials \cite{phi92b,clf+99}.

In an early search of the cluster Terzan 5 \cite{lmd+90}, two radio
pulsars were found, PSR~B1744$-$24A and
PSR~B1744$-$24B. PSR~B1744$-$24A (also known as PSR~J1748$-$2446A) is
clearly associated with the cluster and lies $\sim 30\arcsec$ from its
core.  The pulsar is in a 1.7-hour orbit and is often eclipsed by its
companion star \cite{lmd+90,nt92}.  PSR~B1744$-$24B is rather weak and,
until now, it has not had a phase connected timing solution, so that
its precise position was not known and any association with the
cluster was unclear.

Radio continuum observations at the VLA have revealed the presence of
a number of steep spectrum radio sources, within, or close to, the
core of the cluster, hinting at the possibility of as yet undiscovered
pulsars \cite{fg90}.  Since PSR~J1748$-$2446A is not in the core of
the cluster and we show here that PSR~B1744$-$24B is not associated
with the cluster at all, these two pulsars account for none of the
flux density detected by Fruchter \& Goss (1990)\nocite{fg90}.  This
suggests that there may be further undiscovered pulsars within the
core of the cluster. Any other pulsars may remain undetected as a
result of broadening of the pulses due to dispersion in the receiver
filterbank channels, broadening of the pulses due to acceleration in
tight orbits, very short pulsation period or insufficient sensitivity.
In this paper, we report on a search which was undertaken to try to
resolve these issues and present the one new pulsar that was found,
which we have designated PSR~J1748$-$2446C. We also report on timing
observations of both this pulsar and PSR~B1744$-$24B in order to
establish their positions and associations with the cluster.

\section{Search Observations and Analysis}

The globular cluster Terzan 5 was observed on 1994 April 15 for 1.6 hours
and on 1994 Sept. 14 for 2.8 hours with the Parkes telescope using a 64-MHz
band centred on 1392~MHz. Dual channel cryogenic receivers were used in
conjunction with a 2x256x0.25~MHz channel filter bank.  After detection, the
signals from the two linearly polarised channels were added, one-bit
digitised at a sampling interval of 0.3\,ms and written to magnetic
tape. The beam width was $14\arcmin$ arc and the dispersion smearing per
channel at the cluster dispersion measure of $\sim$240 cm$^{-3}$pc was
0.2~ms, a factor of four better than the earlier search \cite{lmd+90}. For
long-period pulsars, the sensitivity of this search is approximately 40 per
cent better than the previous search.

The data were first processed using standard techniques for
unaccelerated pulsars \cite{mld+96}.  PSR~J1748$-$2446C was found as
an isolated pulsar with a period of approximately 8~ms. To provide
some sensitivity to accelerated pulsars, the two observations were each
split into 20 segments. A Fourier transform was then applied to each
segment and the resulting power spectra were stacked to give a time
versus frequency plot in which accelerated pulsars would show up as
sloping tracks.  Since we were only searching one location and a small
range of dispersion measures (185--260 cm$^{-3}$pc) it was possible to inspect the
stacked spectra by eye.  Despite having sensitivity to millisecond
pulsars in orbits as short as 20 minutes, no further new pulsars were
found this way.

\begin{table*}
\begin{tabular}{llll} 
\multicolumn{4}{c}{Observed Parameters for
Pulsars in or near Terzan~5}\\ 
\hline \hline 
J2000 Name              & J1748$-$2446A & J1748$-$2446C  & J1748$-$2444 \\
B1950 Name           & B1744$-$24A & --  & B1744$-$24B \\
GC Name     & Ter5A & Ter5C & -- \\ 
Right Ascension (J2000) &  $17^{\rm h}48^{\rm m}02\fs255(2)$ 
& $17^{\rm h}48^{\rm m}04\fs54(1)$ & $17^{\rm h}48^{\rm m}48\fs511(5)$ \\ 
Declination (J2000)     &  $-24^{\circ}46\arcmin 36\farcs9(6)$ 
& $-24^{\circ}46\arcmin36\arcsec(4)$ & $-24^{\circ}44\arcmin 43\arcsec(3)$ \\ 
Epoch of Period (MJD)   &       48270.0 & 50958.0  & 51016.0 \\ 
Period (ms)      &       !11.56314838986(3) & !!8.4360953044(1) & 442.838503373(3) \\ 
Period Derivative ($10^{-15}$)   & $-$0.0000340(4) & $-$0.000606(4) & 0.1107(4) \\ 
Dispersion Measure (cm$^{-3}$ pc)       & 242.1(2) & 237(1) & 207.3(2) \\ 
Orbital Period (sec)   &       6535.8240(2) & -- & -- \\ 
Projected Semi-major Axis (lt-s)        &      0.11971(3) & -- & -- \\ 
Eccentricity & 0.0 & -- & -- \\
Epoch of Ascending node (MJD) &  48270.029979(7) & -- & -- \\ 
Data span (days) & 1739 & 1011 & 856 \\
R.M.S. timing residual (ms) & 0.26 & 0.52 &  2.8 \\
Flux Density (mJy)          & 2.5$^{\dag}$ & 0.5(2) & 0.6(2) \\
W$_{50}$ (ms)	            & 0.7$^{\dag}$ & 2.0(5) & 6(1) \\
\hline 
\end{tabular}
\\ 
$^{\dag}$ From Lyne et al. (1990)
\end{table*}

\section{Timing observations and analysis}

Pulse arrival times for the three pulsars in or near Terzan 5 were
obtained on a regular basis since their discovery using the 76-m
Lovell telescope at Jodrell Bank with cryogenic receivers at 606 and
1404\,MHz.  Both hands of circular polarisation were observed using a
$2\times64\times0.125$-MHz filter bank at 606\,MHz and a
$2\times32\times1.0$-MHz filter bank at 1404\,MHz.  After detection,
the signals from the two polarisations were added, filtered, digitised
at appropriate sampling intervals and dedispersed in hardware before
being folded on-line, and written to disk. Observations were typically
of 18 minutes duration.

For each pulsar, a standard pulse template was fitted to the observed
profiles at each frequency to determine the pulse times-of-arrival
(TOAs).  The TOAs, weighted by their individual uncertainties
determined in the fitting process, were analysed with the {\tt TEMPO}
software package (http://pulsar.princeton.edu/tempo), using the DE200
ephemeris of the Jet Propulsion Laboratory
\cite{sta82} and the Blandford \& Teukolsky (1976)\nocite{bt76} timing model
for binary pulsars. The resulting model parameters are summarised in Table 1. 

\section{Discussion}

The new single millisecond pulsar, PSR~J1748$-$2446C (Ter5C), lies
only about $\sim 10\arcsec$ from the cluster centre
($\alpha$(J2000): $17^{\rm h}48^{\rm m}04\fs9$, $\delta$(J2000):
$-24^{\circ}46\arcmin 45\arcsec$ \cite{fg00}) and has a dispersion
measure close to that of PSR~J1748$-$2446A (Ter5A), suggesting with
high probability that it also is associated with the cluster. 

\begin{figure}
\plotone{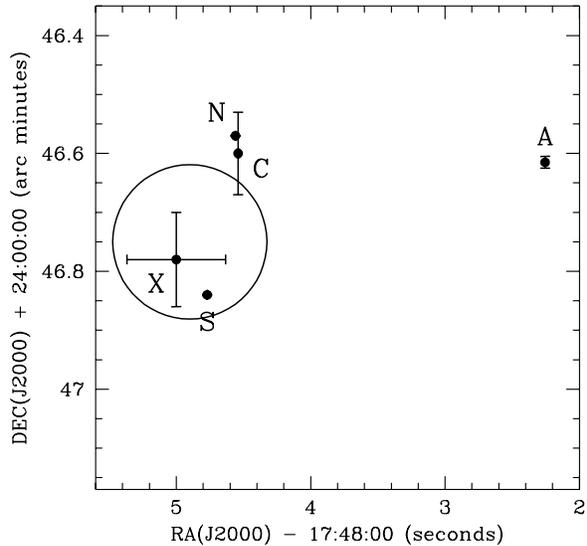}
\caption{Positions of sources associated with the globular cluster Terzan 5.
A: PSR~J1748$-$2446A, C: PSR~J1748$-$2446C, 
circle: cluster core, radius = $7\farcs9$ (Trager et al. 1993), 
N,S: northern and southern radio sources identified by Fruchter \&
Goss (2000), and X: XB 1745--25 (Johnston et al. 1995). }
\end{figure}
\nocite{jvh95} \nocite{tdk93} \nocite{fg00}

The position for PSR~J1748$-$2444 given in Table 1 is $10\arcmin$ from
the cluster centre, indicating that it is unlikely that this pulsar is
associated with the cluster. As Table 1 shows, this pulsar has a
dispersion measure which differs from the values for the other two
pulsars, Ter5A and Ter5C, by about 30 cm$^{-3}$ pc, adding weight to
the conclusion that PSR~J1748$-$2444 is not associated with the
cluster.  The period of 0.443~s and the measured period derivative
indicates that it is an unremarkable pulsar, having a surface magnetic
field of $2.3\times10^{11}$G which is near to the lower end of the
distribution of normal pulsars \cite{tml93}, and a characteristic age
of 63~My.

Figure 1 shows the positions of the two cluster pulsars and other
sources around Terzan 5 relative to the cluster centre.  The two radio
sources (N and S) were found by Fruchter \& Goss (2000)\nocite{fg00}
in a continuum image of the region at around 1400~MHz.  Although the
position of source N agrees within the errors with that of the 8.4-ms
pulsar PSR~J1748$-$2446C, the flux density of the pulsed emission from
this pulsar has a mean value of only $0.5\pm0.2$~mJy.  This is
significantly less than the flux density of source N, 1.5 mJy.
Neither the flux density of the pulsar or of source N vary
significantly between observations, so that such variations cannot
explain the discrepancy.  We conclude therefore that other,
rapidly-rotating pulsars or pulsars in short-period binary orbits may
exist in Terzan 5. Searches with a more sensitive receiving system
having better time and frequency resolution may find them. Johnston,
Verbunt \& Hasinger (1995)\nocite{jvh95} identified a transient X-ray
source close to the cluster core; its positional accuracy is not
sufficient to confirm an identification with the radio source S.

The negative period derivatives of Ter5A and Ter5C indicate that they
lie behind the cluster and are experiencing a gravitational
acceleration towards the cluster core in the manner described by
Phinney (1992)\nocite{phi92b}.  If we assume that the intrinsic period
derivatives are greater than zero and follow the procedures of Phinney
(1992) and Camilo et al. (1999)\nocite{clf+99}, we can place a lower
limit on the central mass density $\rho(0)$ of the cluster. Using the
cluster core radius $r_{c} = 7\farcs9$ \cite{tdk93}, a cluster
distance of 7.6 kpc \cite{jvh95}, and the cluster position of Fruchter
\& Goss (2000)\nocite{fg00}, we find lower limits on $\rho(0)$ of
$0.85 \times 10^5$ and $5.0 \times 10^5$ M$_{\odot}$pc$^{-3}$ for
Ter5A and Ter5C, respectively. We note that the latter is considerably
lower than the value of the density quoted by Webbink (1985)
\nocite{web85} of $24 \times 10^5$ M$_{\odot}$pc$^{-3}$.


\end{document}